\documentclass[runningheads]{llncs}

\usepackage{graphicx}
\usepackage{amsmath}
\usepackage{amssymb} 
\usepackage{mathtools}
\usepackage{color}
\usepackage{xcolor}
\usepackage{xspace}
\usepackage[breaklinks,colorlinks,linkcolor=blue,citecolor=blue,urlcolor=blue]{hyperref}
\usepackage[boxruled]{algorithm2e}
\usepackage{cleveref}
\usepackage{extarrows}
\usepackage{marvosym}

\long\def\comment#1{}

\long\def\com#1{\comment{[{\color{gray} \em #1}]}}

\long\def\cb#1{\comment{ {\bf CB: } [{\color{blue} \em #1}]}}
\long\def\zhq#1{\comment{ {\bf ZHQ: } [{\color{red} \em #1}]}}
\long\def\baf#1{\comment{ {\bf BAF: } [{\color{purple} \em #1}]}}

\newcommand{\eg}{{\em e.g.}}
\newcommand{\ie}{{\em i.e.}}

\com{
\newcommand{\Coin}{Demcoin\xspace}
\newcommand{\coin}{Demcoin\xspace}

\newcommand{\coins}{Demcoins\xspace}

\newcommand{\coinlet}{Demlet\xspace}

\newcommand{\coinlets}{Demlets\xspace}

\newcommand{\popsign}{\mathfrak{d}}

\newcommand{\foresign}{\$}
}

\newcommand{\Coin}{PoPCoin\xspace}
\newcommand{\coin}{PoPCoin\xspace}

\newcommand{\coins}{PoPCoins\xspace}

\newcommand{\coinlet}{PoPlet\xspace}

\newcommand{\coinlets}{PoPlets\xspace}

\newcommand{\popsign}{\mathfrak{p}}

\newcommand{\foresign}{\$}

\begin{document}
\title{Economic Principles of \coin, \\ a Democratic Time-based Cryptocurrency}
%



\author{Haoqian Zhang \and Cristina Basescu \and Bryan Ford}
\institute{Swiss Federal Institute of Technology in Lausanne (EPFL)}




\maketitle              
\begin{abstract}
While democracy is founded on the principle of equal opportunity
to manage our lives and pursue our fortunes,
the forms of money we have inherited from millenia of evolution
has brought us to an unsustainable dead-end of exploding inequality.
\Coin proposes to leverage the unique historical opportunities that
digital cryptocurrencies present for a ``clean-slate'' redesign of money,
in particular around long-term equitability and sustainability,
rather than solely stability, as our primary goals.
We develop and analyze a monetary policy for \coin
that embodies these equitability goals in two basic rules
that maybe summarized as supporting equal opportunity in ``space'' and ``time'':
the first by regularly distributing new money equally to all participants
much like a basic income,
the second by holding the aggregate value of these distributions
to a constant and non-diminishing
portion of total money supply through demurrage.
Through preliminary economic analysis,
we find that these rules in combination yield a unique form of money
with numerous intriguing and promising properties,
such as a quantifiable and provable upper bound on monetary inequality,
a natural ``early adopter's reward'' that could incentivize rapid growth
while tapering off as participation saturates,
resistance to the risk of deflationary spirals,
and migration incentives \emph{opposite}
those created by conventional basic incomes.

\com{	probably not needed for submission
}
\end{abstract}

\section{Introduction}

A well-functioning free market rewards the providers
of valuable products and services,
encourages innovation through competition,
and limits waste by financially starving ventures that fail to produce value.
But today's free markets embody
at least two long-term sustainability problems:
they allow uncontrolled increase in inequality~\cite{piketty17capital},
and they cannot function without constant growth~\cite{kallis12economics,jackson18post}.
For centuries, philosophers and economists have proposed ways in which
the concept and function of money might be improved or redesigned
to be more stable, equitable, and
sustainable~\cite{gesell58natural,keynes65general,jackson09prosperity}.
Permissionless cryptocurrencies,
however,
offer us the unprecedented opportunity
not only to envision on paper a ``clean-slate'' redesign of money,
relatively unconstrained by either the economic status-quo
or risk-averse governments,
but also to \emph{implement} alternative monetary designs
and experiment with them circulating in real-world communities.

\Coin is a cryptocurrency project aiming to prototype and eventually launch
a more democratic, equitable, and sustainable form of money.
Today's ``democratic'' societies and organizations typically attempt
to satisfy the democratic principle of equality only in terms of governance,
via ``one person, one vote'' in decision-making.\footnote{
Many organizations that are often loosely labeled ``democratic''
fail even in this,
including permissionless proof-of-work and proof-of-stake cryptocurrencies,
and open-source communities whose governance is 
dominated by a few core committers.}
\Coin, in contrast, pursues democratic equality
in three dimensions: governance, operation, and economics.
Governance equality means ``one person, one vote''
in decision-making, as usual.
Operational equality in \coin means ``one person, one unit of stake''
in securing consensus and maintaining
a shared history or blockchain~\cite{borge2017proof}.
Finally, economic equality in \coin
means ensuring all participants \emph{equality of opportunity}
to employ money --
and the community resources it represents --
towards personal and collective goods.
This paper focuses on the third objective of economic equality:
the others, while equally important challenging,
we leave for other work to address.

\com{
\Coin departs from traditional currencies and cryptocurrencies alike
by taking \emph{people}, rather than \emph{things},
as its basic foundation for monetary value.
}

Motivated by supporting sustainable and equitable economic opportunity
while retaining capitalistic rewards for valued work and innovation,
\coin regularly mints and distributes new money to all
\emph{real human} participants.
The basic goal of \coin's monetary policy is to ensure
that these ``basic income'' distributions~\baf{citations}
provide all participants a baseline of economic opportunity
that is equitable, continuous,
and unvarying in both space and time.
Equality over ``space'' -- \ie, population --
means guaranteeing each participant
an equal share in each distribution of new money.
Equality over time means the value of each basic income distribution
represents an equal and constant proportion
of the community's total monetary resources.
Most critically, the basic income's proportionate value and utility
must not diminish from one month, year, or generation to the next.
\com{
Embodying these two constraints in a cryptocurrency's monetary policy
leads to intriguing and promising properties.
}

This paper's first main contribution is
a preliminary ``long view'' economic model and analysis
of what a sustainable, democratically egalitarian form of money might look like,
if it were eventually to become widely-adopted
as the predominant currency within a community of sufficient critical mass.
Our analysis adapts existing tools from economic theory
to develop \coin's monetary policy and identify several interesting properties.
First,
in line with pre-cryptocurrency ideas such as
Gesell's \emph{freigeld}~\cite{gesell58natural} and 
universal basic incomes~\cite{parijs17basic,standing17basic},
\coin decouples broad economic growth from debt
by giving all participants a regular supply of debt-free money.
Second, \coin's design imposes a readily-calculable upper bound on inequality
in the distribution of \coin among participants after each minting,
thereby ensuring a form of sustainability in terms of basic financial inclusion,
contrasting with classic currencies where inequality
can increase without bound.
Third, while increased real growth
can create monetary deflation by increasing demand,
\coin appears to mitigate the classic risk of ``deflationary spirals'' --
where high demand yields higher real interest rates and hence borrowing costs,
making money even more scarce in a positive feedback loop.
Because \coin's basic income is not debt-based,
higher demand on \coin raises the basic income's real value
without affecting its broad availability to participants,
and may counteract deflation by giving all participants
both opportunity and purchasing power to spend or invest.

While this long view is promising,
real currencies are not isolated but inhabit a larger economic ecosystem.
This paper's second main contribution
is a preliminary exploration of several intriguing properties we may expect of
a \emph{permissionless cryptocurrency} embodying \coin's monetary policy.
First, a preliminary exchange rate analysis
suggests that we may expect \coin to increase gradually in value over time
with respect to an inflationary fiat currency like USD,
assuming that both the size and spending behavior of the \coin community
is stable in certain respects.
We don't actually expect either of these factors to be stable in practice,
however, especially while a \coin community is small and rapidly-evolving,
leading to a second key observation.
Because basic income represents a fixed fraction of total \coin supply
divided by participant population in \emph{that} minting,
earlier participants in a growing community
receive a larger fraction of total supply in earlier mintings.
This effect might serve as a natural ``early adopter's reward'' --
and an incentive for participants to promote \coin and grow the community --
that automatically tapers off as participation saturates and stabilizes.
Third, while we may expect speculative trading and ``HODLing'' of \coin
to yield wild exchange-rate swings as with other cryptocurrencies,
\coin continually ``taxes'' speculative holdings
and redistributes value to participants' basic incomes,
which may both disincentivize too much speculative holding
and reward participants for weathering speculative storms.
Nevertheless, we prove that a rational participant may 
save some of his income to increase his future utility, 
and the rich would pay more demurrage fee than the poor, 
even though they are subject to the same mechanism and global demurrage rate.
Fourth, because \coin represents a permissionless and borderless community,
its basic income floats to some single global value versus other currencies,
rather than being defined by policy within a jurisdiction
as in a conventional basic income\baf{citations}.
\coin's basic income will therefore buy more and feel more useful
in poor countries with low cost-of-living than in rich ones,
it will gradually redistribute wealth from the latter to the former,
and any ``migration incentive'' it creates will be
from rich countries to poor:
opposite the poor-to-rich migration incentives
that a conventional UBI would contribute to.

This initial development of \coin has many limitations, of course.
It focuses only on monetary policy,
leaving operational and governance issues out of scope.
Our preliminary economic analysis, detailed in the appendices,
makes many simplifying assumptions that may prove unrealistic,
and our formal model currently covers
only a subset of the interesting properties of \coin
that we identify and explore intuitively.
Further,
because \coin as developed here effectively ``taxes'' only money
and not non-monetary wealth such as real estate and investments,
it inherently incentivizes spending over holding.
This may be desirable to stimulate economic activity
as Gesell proposed~\cite{gesell58natural} --
but it may also make other currencies and non-monetary wealth
more attractive as investements,
limiting the total value we can expect \coin to acquire,
and similarly limiting its potential to address inequality in general
across all forms of wealth.
Indeed we expect, and accept,
that \coin may not appeal much to economically ``greedy'' users,
but mainly to those motivated more by social, political,
or sustainability goals.\footnote{
	Such goals may of course be considered economically rational
	in terms of ``enlightened self-interest'' as opposed to greed.
}
We leave these limitations to be addressed in future work.

\baf{ work-in-progress continuing...  }

\baf{
address real versus nominal economic indicators.
}

\baf{
somewhere briefly cite and relate to the prior PoP papers,
in third-person for anonymization...

\coin's ``proof-of-personhood'' basis for
stake and governance~\cite{borge2017proof,ford08nyms}
and the philosophical principles motivating
its monetary design~\cite{ford19money-anon},
this paper's contribution is ...
}

\baf{move to section 2?

One central use of money is
to trade for a person's time via paid labor or services.
Physical reality continually supplies all living people with time
at an inherently-equal rate,
ignoring relativistic space travel for now.
On this basis, the democratic principle of equal opportunity leads \coin
to adopts the time-banking tradition~\baf{time-banking citations}
of supplying each person a constant supply of new currency at an equal rate.
\Coin does not presume that every person's time is equally valuable to others,
however:
if one \coin represents one hour of unspecialized work that anyone can perform,
a specialist in high demand may charge several, or many, \coins per hour.
}

\baf{
The critical concepts to capture:

- long-term fairness and sustainability as a primary goal
- founded on peoples' time rather than things as the foundation of value
- equal opportunity both over population and across time/generations
- embodies a form of UBI, but with different incentives
- offers free-market incentives, ``equal opportunity to become unequal'',
	but with a designed-in upper bound on inequality
- avoid the "grow or die" problem for sustainability
- a different kind of stability: based on commerce behavior rather than linking to a legacy currency or active feedback?

}

\com{
\emph{Note for reviewing:
some of the ideas in this paper first appeared online
in an earlier preprint by the same authors,
which is neither published nor under submission
and contained no economic analysis details.
An anonymized version of that preprint is available for reference at
\url{https://drive.google.com/file/d/1Xeh3ekFUe2-h2Cj66SHt2Rj1HTEkU_jY/view?usp=sharing}.
}
}

\section{Monetary Principles and Design of \Coin}
\label{sec:meta}

\baf{will need at least one sentence of lead-in summary here.}
\zhq{Perhaps we need some sentences to link with the modeling part. Why do we model? What is the beauty of modeling? What are the conclusions from the modeling that is not so easy to illustrate using text?}

\subsection{Democracy, Money, and the Principle of Equal Opportunity}
\label{sec:meta:equality}

Democracy,
or literally \emph{rule by the people}~\cite{fleck2006origins},
has no single definition but embodies widely-held principles.
The Council of Europe boils democracy down to two key principles:
\emph{individual autonomy} --
that ``People should be able to control their own lives (within reason)'' --
and \emph{equality} --
that ``everyone should have the same opportunity to influence the decisions
that affect people in society''~\cite{coe-democracy}.
Political philosopher Robert Dahl defines criteria essential to democracy,
among them equal opportunity
to obtain ``enlightened understanding'' of the issues,
to control the agenda,
to participate in discussions,
and to vote on decisions~\cite{dahl89democracy}.
We normally interpret ``equality'' only as \emph{political equality}
in self-governance.

Among the ``decisions that affect people in society'', however,
classical economics carves out a huge swath --
namely, almost all decisions on allocating society's resources --
in which \emph{inequality} rules.
We have mostly exempted money from the democratic principle of equality:
those with more money can spend proportionally more on what they like,
hire more labor to help them, invest more in ventures they support, etc.
With inequality exploding~\cite{piketty17capital},
leaving 90\% of incomes stagnant as \$2.5 trillion was transferred
to the top 1\% since 1975~\cite{price20trends},
current trends toward \emph{unlimited inequality}
represent a clearly unsustainable path.

Furthermore, economics and governance are inseparable in practice:
``money is power.''
More money buys more influence~\cite{gilens14theories} --
whether via advertising, lobbying,
or online bot farming~\cite{broniatowski18weaponized}.
For the growing global ``precariat''~\cite{standing16precariat}
struggling to survive on stagnant incomes from multiple uncertain sources,
finding the time even just to vote --
let alone fulfill Dahl's democratic criteria of ``effective participation''
and ``control of the agenda'' based on ``enlightened understanding'' --
feels increasingly like a distant luxury only the rich can afford.
In numerous ways, economic inequality corrodes political equality
and undermines democracy.

Irving Fisher noted the unsustainability
of uncontrolled economic inequality,
and its corrosion of political equality,
after the conclusion of World War I
in his 1919 annual address as president of the American Economic Association:

\begin{quote}
Our society will always remain an unstable and explosive compound as long as
political power is vested in the masses and economic power in the classes. In
the end one of these powers will rule. Either the plutocracy will buy up the
democracy or the democracy will vote away the plutocracy. In the meantime the
corrupt politician will thrive as a concealed broker
between the two.~\cite{fisher19economists}
\end{quote}

Political and economic philosophers alike often support
the principles of inclusion and \emph{equal opportunity}.
Even capitalist economics generally presumes that participants compete
on a fair and ``level playing field''
even if \emph{outcomes} may -- and arguably should -- be highly unequal.
Among the opportunities most people want
are the opportunities to earn economic rewards for hard work,
innovation, or wise investment.
For these purposes,
we cannot realistically pretend that everyone is equal
in either abilities or motivation.
But if we accept that allowing (equal) opportunity to earn rewards
necessitates allowing inequality in economic outcomes,
this does not mean we must or should accept \emph{unlimited} inequality.

Classical monetary policy is driven primarily by stability concerns:
particularly stable prices, to protect money's functions
as a unit of account and a store of value,
and a stable money supply to drive commerce~\cite{keynes65general}.
Even at this modest goal of maintaining a stable ``status quo,''
however,
classical economics fails miserably,
yielding frequent ``boom-and-bust'' cycles that show
no signs of abating~\cite{reinhart11this}.
But as Bitcoin~\cite{nakamoto2019bitcoin}
and the countless cryptocurrencies it inspired have underlined,
money is not only a social good but a technology that can be \emph{designed},
and some designs will serve us better than others.
We now have the opportunity not only to rethink
but also to implement and deploy
new forms of money without anyone's permission.
Money can now be created electronically by ordinary individuals,
not just by banks.
And even wildly-unstable digital currencies can capture tremendous interest
and enter widespread use.

The central idea motivating \coin
is the radical question of whether in focusing single-mindedly on stability,
classical economics got its basic priorities wrong?
Stability is great when we can get it,
but a stable march towards global economic (and environmental) destruction
is eventually just as disastrous as an unstable march to the same end.
Could we design, implement, and deploy a form of money
that instead pursues sustainability and equal opportunity as its primary goals,
with stability as a still-desirable but subsidiary objective?

\com{
we want people to have opportunities to obtain economic rewards
for hard work and wise investment,
and accept the unequal outcomes that implies.
In order to achieve this,
we must somehow divide the economic power that money offers
into a portion representing a \emph{baseline supply}
to provide equal opportunity,
and a portion representing \emph{rewards}
from past effort and investments.
A natural way to embody this principle in monetary policy --
making room for both equal opportunity and unequal outcome --
is to divide the supply of money into
an equally-distributed ``baseline supply''
to provide the ``level playfield''
and a portion representing past rewards.
}

\com{
In democratic money, everyone should obtain equal portions of the newly minted
money throughout the time, just like that every eligible voter should have
equal voting power in every election, at least in the early phase. Democratic
money should also prevent the effect of ``the rich get richer'', that the
success of past generations dominate future generations' opportunities becoming
even richer and transfer their wealth to their lucky heirs, as summarized in
the work of Thomas Piketty\cite{capital}, to ensure the equality over
generations.
}

\subsection{Towards Sustainable Equal Opportunity in Space and Time}
\label{sec:meta:sustainable}

The design space of cryptocurrencies and monetary policies
is clearly rich and infinite,
so we cannot expect to find any unique or best ``Answer''
to the above challenge.
But could we find some relatively simple monetary policy
that plausibly achieves these goals under arguably-realistic assumptions --
ideally a policy we can encode into a few simple rules
that a cryptocurrency can enforce automatically?
Classical economic theory calls for constant guidance from central banks
to maintain a semblance of economic stability.
Could we find a ``hands-free'' rule-set with the potential
to avoid at least the most destructive instabilities --
namely positive feedback loops such as overheating or deflation spirals --
while ensuring equal opportunity in some formally definable fashion?
In \coin we develop two simple rules that appear
particularly promising in combination.

The first rule is that all participants regularly receive an equal supply
of newly-minted money as a baseline foundation for economic opportunity,
which we refer to as equal opportunity in ``space".
This rule relates closely to the increasingly-popular idea
of basic income~\cite{parijs17basic,standing17basic},
but expressed in \emph{monetary} rather than a purely social policy -- 
a distinction that yields important differences
we explore later in \cref{sec:economic}.
The philosophical grounds for an \emph{equal supply} of basic income
is clearly to support equal opportunity,
a justification often debatably ascribed
to John Locke~\cite{locke1689second,layman11locke,moseley11lockean}.
Our justification for a \emph{regular} supply is
to ensure that support for equal opportunity remains inclusively available
for life
in the face of personal losses from risk-taking, accidents, disasters, etc.,
as discussed further in \cref{sec:meta:regular}.

The second and less-precedented rule underlying \coin is that
the portion of total money supply distributed equally to all participants
must be constant at each distribution,
in particular not diminishing with time, 
summarized as equal opportunity in ``time".
Basic income proposals typically rely on a policy decision
to choose some ``appropriate'' value that somehow balances
standard-of-living expectations against fiscal budgeting constraints.
But the ``right'' balance between expectations and affordability
is infinitely debatable and subject to change frequently
with public and government mood.
Moreover, any chosen value denominated in an inflationary fiat currency
will diminish in real value and effectiveness in time,
just as minimum wage protections have eroded~\cite{cooper19low-wage}.
\Coin introduces the more radical proposition
that we simply peg the value of each distribution
to a constant fraction of total money supply,
chosen and justified on some less-fluid basis,
such as the lifetime an average person has to enjoy or re-invest rewards,
as discussed in \cref{sec:meta:demurrage}

These rules work together toward ensuring
that the economic opportunity offered by regular distributions of new money
is egalitarian in both ``space and time'':
\ie, individually between the participants in any given distribution,
and collectively between earlier and later distributions.
Instead of attempting to support some particular standard of living,
\coin attempts to ensure that all money distributions
are \emph{fair} and \emph{proportionate}:
both individually among the participants in today's distribution,
and collectively with respect to the portion of monetary wealth
similarly distributed in prior months, years, or generations.

This combination of rules supports economic sustainability in two respects.
First, 
\coin's first rule ensures that all participants have
an equal and inalienable source of \emph{debt-free} income,
which could break the economy's reliance on constant growth
to achieve broad-based increases in living standards
as discussed in \cref{sec:meta:growth}.
Second,
\coin's second rule ensures that one's opportunities today are not dominated
by economic wins and losses of the past,
and that inequality cannot increase without bound
as we explore later in \cref{sec:inequality}.
While \coin directly addresses only economic
and not environmental sustainability
as some proposals do~\cite{helbing20futurict},
limiting growth dependance and inequality could reduce pressures
that often result in environmentally unsustainable practices and policies.
\baf{find a good citation or two}

\com{ discuss somewhere the intergenerational redistribution problem
due to government debt}

\com{ discuss somewhere the illustrative idea of a government
whose constitution prevents new laws from overriding old ones,
leaving later generations' policy decision space successively narrower. }

\com{ real estate ``Stand on Zanzibar'' illustration
unsustainability of unlimitied equality given finite resources 
}

\subsection{Time is Money: The Semantic Meaning of a \Coin}
\label{sec:meta:time}

A unit of fiat currency today generally represents an arbitrary unit of value,
whose nominal value has no meaning except in relation to perceived real value
and to other currencies as it floats through supply and demand.
One dollar doesn't ``mean'' anything.
\Coin, in contrast, builds on an idea pioneered by
time-based currencies~\cite{miller08teruko,cahn92time,jacob12social}.
Since money is so commonly used to trade peoples' time
in providing labor and services,
why not \emph{define} the a currency's value in terms of a person's time?
Time is an inherently-stable reference point,
whose advance we can quantify precisely in terms of other physical phenomena,
as atomic clocks do.
Time is also inclusive and democratically egalitarian,
in that everyone living inherently receives a constant ``supply'' of time
at the same rate as others --
ignoring space travelers at relativistic velocities for now.

Imagine a purely-fictional world in which
all people spend
eight hours each day supplying services to someone else,
eight hours per day consuming the services of others,
and eight hours per day sleeping.
Further suppose that all work consists solely of unspecialized services
that anyone can perform,
like sitting with someone to keep them company,
so that one hour of anyone's time is worth exactly as much
as an hour of anyone else's.\footnote{
	Some time-based currencies ask participants
	to trade and bank units of time literally,
	at a fixed ``exchange rate'' regardless of the service.
	This approach has unsurprisingly proven most successful
	in service areas such as child- and elderly-care,
	where trust and community spirit may readily be seen as
	more important than specialized skills.
}
Then the only need the inhabitants have for ``money''
is to negotiate \emph{which} hours each person spends
working (and for whom), consuming services (from whom), or sleeping.
In this fictional world,
one \coin would represent exactly one hour of anyone's time.

We make no pretence that this fictional world models reality,
but treat it merely as an ``ideal reference''
against which we may treat complex reality
as a (large) set of adjustments,
the cumulative effects of which
we let supply and demand reveal dynamically.
Since not everyone's time is equally valuable, for example,
a specialist whose time is five times more sought-after
than that of a fully-unspecialized worker
would find herself able to charge five \coins per hour,
all other factors corresponding to the ideal reference.
We similarly expect the trade value of a real \coin
to deviate from the ideal for myriad other reasons:
\eg, people like to work only five days per week and take holidays;
participation changes due to people adopting or leaving \coin;
people also use other currencies competing with \coin;
investors artificially increase \coin scarcity by HODLing it
and thereby keeping it out of commercial circulation;
usage changes and shocks in other economically-linked currencies
affect \coin indirectly;
ad infinitum.

While accepting that reality is fluid
and far too complex to analyze all the factors,
nevertheless the ambition is that one \coin
should always conceptually ``mean'' something with respect to peoples' time,
should do so equitably,
and should mean the same thing in a decade or a century as it does today.

\subsection{The \Coin Algorithm for Monetary Policy}
\label{sec:meta:algorithm}

\begin{algorithm}[t!]
\caption{Pseudocode representation of \coin monetary policy}
\label{alg:popcoin}

\SetCommentSty{textnormal}
\DontPrintSemicolon
\SetKwInOut{Input}{input}
\SetKwFor{For}{for}{}{end}

\Input{$B$, the number of \coins issued to each participant per minting}
\Input{$\alpha$, the fraction of total \coin supply redistributed per minting}
\BlankLine
$N_0 \leftarrow$ initial number of participants at launch\;
\For(\tcp*[f]{one minting per time period}){$t = 1$ \KwTo $\infty$}{
	$N_t \leftarrow$ number of participants at time $t$
		\tcp*{take new participation census}
	scale wallet balances by $N_t / N_{t-1}$
		\tcp*{adjust for participation changes}
	scale wallet balances by $1 - \alpha$
		\tcp*{apply demurrage to existing coins}
	issue each participant $B$ \coins
		\tcp*{distribute new basic income}
}
\end{algorithm}

We can now specify the \coin monetary policy concisely
in terms of the pseudocode in \cref{alg:popcoin}.
This algorithm is concerned only with what happens at each regular minting,
and assumes that wallets and normal trading between mintings
are handled by standard (\eg, Bitcoin-like) transaction processing.
In brief, at each minting the algorithm
(a) determines the new number of participants at time $t$,
(b) redenominates the currency to account for participation changes,
(c) applies demurrage to current balances to keep total supply constant, and
(d) issues a constant $B$ new \coins to each participant.
We briefly unpack and informally justify each step below.

Because we want one \coin to represent one hour of unspecialized work
in the ideal reference world above,
and each of the $N_t$ participants at time $t$
have an inherent supply of $B$ work-hours of time per minting period,
total \coin supply must depend on -- and be proportional to -- participation.
To account for participation being non-constant
in a real permissionless cryptocurrency,
\cref{alg:popcoin} effectively \emph{redenominates} the currency
in step (b) whenever participation changes.
That is, the algorithm simply scales all existing wallet balances
by the factor $N_t/N_{t-1}$
to convert the last time period's currency --
appropriate for $N_{t-1}$ participants --
into a ``new currency''
appropriate for $N_t$ participants.

To ensure that \cref{alg:popcoin} can issue each participant
a fixed number $B$ of new \coins per minting in step (d),
while also ensuring that the total value of all newly-minted currency
represents a fixed fraction $\alpha$ of the currency's total supply
as discussed in \cref{sec:meta:sustainable},
we must \emph{demurrage} all existing wallet balances in step (c)
by the factor of $1 - \alpha$.
This demurrage ensures that total \coin supply at time $t$
asymptotically approaches but never exceeds $B N_t / \alpha$.

Redenominating a conventional currency with printed banknotes
is of course an expensive process
typically done only rarely
after periods of inflation~\cite{mosley05dropping}.
Demurrage is similarly nontrivial with printed banknotes --
a purpose for which Gesell~\cite{gesell58natural}
invented the clever idea of \emph{stamp scrip},
where the holder of a banknote must purchase and periodically affix
stamps weekly in order to keep the banknote
valid~\cite{fischer33stamp,champ08stamp}.
Redenomination and demurrage are
straightforward in principle for a Bitcoin-like cryptocurrency, however,
where all wallet balances reside on a shared ledger.
We may worry that currency users may be confused and concerned
on seeing their nominal wallet balances change periodically --
but this already happens with conventional bank accounts
when maintenance fees are charged, interest is deposited, etc.
Further, the implementation-efficiency issue
of regularly updating all wallet balances
is readily addressed by internally denominating wallet balances
in an inflationary and participation-independent ``hidden currency''
like \emph{\coinlets} as discussed in \cref{sec:poplets}.

Simplistically assuming one minting per year,
we would set $B = 365.25 \times 8 = 2922$
to reflect the ideal reference model above in which one \coin
represents eight hours of unspecialized work
per day.\footnote{``
	More frequent mintings, likely more practical,
	just require adjusting the constants.
}
The fraction of total supply demurraged and redistributed
at each minting, $\alpha$,
is similarly somewhat arbitrary
but might reasonably be set to 2\%,
giving \coin a 50-year ``tenure''
corresponding to around a modern human working lifespan
as discussed in \cref{sec:meta:demurrage}.

\section{A Basic Model for \coin with Three Principles}

We now introduce an economic model
built on methods typically used in monetary policy by central banks
and other cryptocurrencies.
We then express the principles of \coin in this model:
fixed basic income, equality over population and equality over total supply.
Finally, we derive \coin's monetary policy from these principles alone
and argue that \coin's monetary policy is both necessary and sufficient
to fulfill them.

\subsection{Economic Modeling Assumptions}

\begin{table}[t]
\centering
\begin{small}
\begin{tabular}{ |c|c| } 
 \hline
 Symbol & Meaning \\
 \hline\hline
 $M$ & Money supply. \\ 
 \hline
 $R$ or $r$ & Interest rate. \\
 \hline
 $B$ or $b$ & Basic income expressed in amount of \coin given to each participant. \\
 \hline
 $D_t$ & Amount of total basic income distributed in period of $t$. \\
 \hline
 $x_i$ & Account balance of person $i$, expressed in \coin. \\
 \hline
 $N$ & Population size.  \\
 \hline
 $\alpha$ & The ratio of newly issued \coin to total supply.\\
 \hline
\end{tabular}
\end{small}
\label{table:notation}
\newline
\caption{Notation. The uppercase letters represent macroeconomic variables, whereas the lowercase ones represent microeconomic variables.}
\end{table}

Monetary policies, whether directed by central banks or encoded in cryptocurrency code, 
are generally built on two methods to control the money supply:
adjusting interest rates and directly distributing money.
\footnote{
	Interest-rate driven monetary policy used to be the most effective
	standard tool used by central banks.
	As interest rates plummeted to zero,  however,
	central banks have had to adjust money supply more directly,
	either through quantitative easing or 
	by directly handing ``helicopter money''
	to spenders\cite{dalio2018principles}.
	Cryptocurrencies like Bitcoin and Ethereum, in contrast,
	generally do not have the concept of an interest rate:
	They solely rely on block rewards as their mechanism
	to inject money and increase the money supply.
}
Our economic model incorporates both of these methods.
Table~\ref{table:notation} summarizes the notation we use subsequently
throughout this paper.

\subsubsection{On a Micro Level} we assume a world
with $N_t$ participants during a time period $t$,
where $t\in\mathbb{N}$. 
We denote by $x^i_{t}$ the balance at the end of period $t$ 
for any participant $i$, $i=1, 2, \dots, N_t$.
Naturally, $x^i_{t}$ carries into the next period $t+1$, 
as savings that potentially yield interest. 
Our model assumes the interest on savings as part of the income of any participant
during each period of time. 
We further denote by $r_{t}^i$ the interest rate on savings 
of participant $i$ from period $t-1$.
\footnote{The interest rate can be zero, 
\ie~no interest; positive, 
\ie~a currency holder can earn positive capital income from his saving; or negative, \ie~a currency holder needs to pay for keeping his saving. 
Although it is common to assume for a positive interest rate, 
a zero or negative rate is not unseen.
At the time of writing
the central banks of Japan, Denmark, and Switzerland 
are implementing negative interest rates\cite{interest-rates}.}

In addition, a participant $i$ also generates 
the following two types of income 
at the beginning of any time period $t$: 
(a) the basic income $b^i_{t}$
distributed to each participant,
and (b) the earned income, denoted by $in^i_{t}$. 
Participant $i$'s balance may accumulate over time \ie 

\begin{equation}
\label{incomeAndSpending}
     x^i_{t}=x^i_{t-1}(1+r^i_{t})+b^i_{t}+in^i_{t}-out^i_{t}\,,
\end{equation}
where $out^i_{t}$ denotes the expenditure
of participant $i$ in period $t$. 

\subsubsection{On a Macro Level} 
we assume that both basic income and positive interest income
are freshly-minted,
\ie~ created by the monetary system purely numerically --
instead of relying on any social entity such as a government or philanthropist, 
or by compelling one into debt.
These are the only two methods to mint new coins.
Similarly, we assume that the negative interest simply ``disappears'', 
which leads to a reduction of the total currency supply. 
Therefore, the net change of the total supply 
becomes the sum of all participants' basic income and interest in the currency:

\begin{equation}\nonumber
    \sum_{i=1}^{N_t} x^i_{t} - \sum_{i=1}^{N_t} x^i_{t-1}=\sum_{i=1}^{N_t} x^i_{t-1}r^i_{t}+\sum_{i=1}^{N_t} b^i_{t}\,,
\end{equation}
or -- on a macro level -- we write:
\begin{equation}\label{eqn.supply}
    M_{t} - M_{t-1} = R_{t}M_{t-1} + D_{t}\,,
\end{equation}
where $M_{t}$ denotes the total supply at the end of period $t$, 
and $D_{t}$ denotes the sum of total basic income distributed in period $t$.
The aggregate interest rate for all participants, unique in any period $t$, 
we denote by $R_{t}$.

\subsection{The Principles of \coin}
\label{principles}
The principles of achieving equal opportunity in both ``space and time'' from
\cref{sec:meta:sustainable} 
define the key policy constraints defining \coin.
We therefore abstract these principles into 
the following three mathematical equations 
representing \coin's fundamental principles:

\begin{subequations}

\subsubsection{(I) The Principle of Fixed Basic Income.} 
Because \coin is a time-based currency,
each participant periodically receives the fixed nominal amount $B$
of basic income
as introduced in Section~\ref{sec:meta:algorithm} above,
\begin{equation*}
     B^i :=B^i_{t-1} = B^i_{t}\,,
\end{equation*}
where $B^i$ denotes the basic income of participant $i$. 

\subsubsection{(II) The Principle of Equality over Population.} 
Because basic income is intended to support equal opportunity in \coin,
its amount must be the same across all participants.
We denote this universal amount as $B$, \ie,

\begin{equation}\nonumber
     B:=B^i= B^j \,,
\end{equation}
where $i$ and $j$ denotes two different participants.
We therefore obtain the total amount of newly issued \coins as:
\begin{equation}
\label{distribution}
    D_{t}=N_t B\,.
\end{equation}

\subsubsection{(III) The Principle of Equality over Total Supply.} 
\coin achieves equality across time and generations
by setting the total amount of newly issued \coin 
to be a fixed proportion of the existing \coin supply, \ie,

\begin{equation}
\label{relationMDp}
    \frac{D_{t}}{M_{t}} = \alpha\,. 
\end{equation}
with $\alpha$ a constant system parameter. \Cref{sec:meta:demurrage} discusses how we might choose the value of $\alpha$.

\end{subequations}

\subsection{Deriving the \coin Monetary Policy From Its Principles}

\label{sec:mechansim}
Adopting all Principles (I)$\sim$(III) uniquely 
determines the monetary mechanism of \coin; 
any change to the mechanism would refute at least one of its principles.
In the following, we derive the mechanism of \coin step by step 
from a strawman case with a constant population size and money injected through basic income, but no interest rate, to a simplified scenario that allows for interest rate, and finally 
to a general situation with varying population size.

\subsubsection{Strawman: Fixed Population, Basic Income and Zero Interest.}
With a fixed number of participants, 
Principles (I) and (II) set
the total basic income distributed in each period to a constant,
according to \cref{distribution}.
Under a zero interest rate setting over all time periods
like Bitcoin and Ethereum, 
the total amount of \coins would grow linearly: 

\begin{equation*}
    M_{t-1}+D_{t}=M_{t} .
\end{equation*}

This would violate Principle (III), however,
in that the issue-to-supply ratio (\cref{relationMDp})
would decrease over time as a result.

\subsubsection{Simple Case: \coin with Fixed Population.}
By this reasoning, with constant participation,
Principle (III) can be maintained only
under a negative interest rate, \ie~ $R_{t}=-\alpha$.

We substitute $D_{t}$ in \cref{eqn.supply}
by its form in \cref{relationMDp}. 
Under the negative interest rate $-\alpha$, 
the total currency supply over time becomes:
\begin{equation}
\label{eqn.supply1}
    M_{t} = M_{t-1}(1-\alpha)+D_{t}\,.
\end{equation}

A negative interest rate or \emph{demurrage}
devalues existing coins gradually over time~\cite{lietaer2006community}.
In this case, the demurrage rate is exactly $\alpha$.

\subsubsection{General Case: \coin with Population Changes.}

Now consider the general case with population changes.
We denote by $n_t$ the population growth rate in period $t$,
such that $N_t = (1+n_t)N_{t-1}$. 
By \cref{distribution},
the total amount of newly issued \coin, or $D_t$, grows at the same rate: 

\begin{equation}
\label{eqn.growth}
    D_{t} = D_{t-1} (1+n_t)\,.
\end{equation}

Solving the system of \cref{eqn.supply}, \cref{relationMDp} and \cref{eqn.growth}
for the interest rate, 
we obtain $R_{t} = (1+n_{t})(1-\alpha) -1$.
Total currency supply now takes the following form:
\begin{equation}
\label{OverTime}
    M_{t} = M_{t-1}(1+n_{t})(1-\alpha)+D_{t}\,.
\end{equation}

This matches our \cref{alg:popcoin} that existing \coins scaled 
by the factor of $(1+n_{t})(1-\alpha)$ with 
new basic income distributed to every participant.

\subsubsection{\coin Supply Grows Proportionally to the Population Size}
We denote $\mu_{t}$ as the growth rate of \coin, we have:

\begin{equation}
\label{changerate}
    \mu_{t}= n_{t} .
\end{equation}

This equation can be clearly presented from the relationship 
between the currency supply and the population size, 
implied by the combination of \cref{distribution} and \cref{relationMDp}, \ie,

\begin{equation}
\label{participants}
    M_{t} = \frac{1}{\alpha}BN_t\,,
\end{equation}
and it is not hard to verify that it is consistent with \coin's monetary policy shown in \cref{OverTime}.

\section{Preliminary Economic Analysis of \coin}
\label{sec:economic}

Preliminary analysis of the above model
leads us to a number of interesting observations about \coin,
as detailed in the appendices and summarized here.
These analyses of course make many simplifying assumptions
and cannot hope to model all the complex factors relevant in reality,
but they allow us to tease apart some broad effects and trends.

\cb{We analyze \coin using the standard assumptions made in economic literature by Feenstra and Taylor~\cite{feenstra2017international}. We aren't the first to make these assumptions and, while extensively debated in economic literature, they're still accepted as a reasonable first step to model the world.}

\subsubsection{Bounded Inequality.}
\Coin's most important property from a sustainability perspective
is that it establishes an upper bound on inequality,
at least in terms of monetary wealth denominated in \coin.
\Cref{sec:inequality} shows that after each basic income minting,
\coin ensures an upper bound in three inequality metrics:
Gini coefficient,
variance across all participants' balances,
and ratio between any two balances.
Limiting inequality in monetary wealth alone this way would not, of course,
necessarily bound inequality in general across \emph{all} forms of wealth,
even in a hypothetical population that used only \coin as money.
Nevertheless, to the degree that having access to money
with which to engage in commerce and seek to improve one's fortune
is a key element of economic opportunity in practice,
\coin might ensure that this social good and driver of opportunity at least
cannot become too unevenly divided over time.

\subsubsection{Adoption Incentives.}
When participation grows more quickly than the demurrage rate of $\alpha$,
\coin's monetary policy offers a natural ``reward'' to early adopters --
along with an incentive for early adopters
to promote \coin and further increase adoption.
Suppose the number of participants doubles in some period $t$,
for example, so $N_t = 2N_{t-1}$.
Then the basic income $B$ that an early adopter received at time $t-1$ will,
after redenomination for the population change in \cref{alg:popcoin},
have a nominal value of $2B$ at time $t$ before demurrage.
The early adopter's saved basic income from time $t-1$, therefore,
is effectively worth almost two basic incomes at time $t$.
This does not mean that the \emph{real} value of these savings
necessarily doubles correspondingly, of course.
But if the new adopters put the currency
in active use and circulation similarly to the existing users,
thereby growing the real \coin economy roughly proportionally as well,
then the early-adoption reward will also be meaningful in real value.

These effects will naturally create speculation incentives
while population and/or currency usage is rapidly evolving.
Rational investors who correctly predict at time $t-1$
that participation will double by $t$, for example,
may be willing to buy \coin from other participants at $t-1$
for close to twice what they expect it to be worth at $t$,
precisely to take advantage of the early adopter's reward.
Speculation may well create wild swings in \coin's trade value,
just as with other cryptocurrencies.
But recall that \coin's primary goal
is \emph{long-term} fairness and equitability,
with stability only a subsidiary goal,
as discussed in \cref{sec:meta:equality}.
As \coin gradually saturates some population of receptive users,
the early-adoption reward tapers off and disappears as participation stabilizes,
leaving demurrage as an incentive to spend rather than hold \coin. 
We leave the mathematical definition in \cref{sec:population}.

\subsubsection{Exchange Rate Analysis.}
A preliminary exchange rate analysis in \cref{sec:exchange},
both long-run and short-run,
indicates that \coin would gradually appreciate
with respect to inflationary fiat currencies
assuming other factors remain stable.
For the long-run analysis,
we assume that price is flexible
and \textit{Purchasing Power Parity} holds~\cite{a19901}.
The short-run analysis yields a similar conclusion,
together with the expected 
\textit{exchange rate overshooting} phenomenon~\cite{feenstra2017international},
under the assumption that 
prices are sticky and \textit{Uncovered Interest Parity} holds~\cite{isard2006uncovered}.
In summary, the fact that \coin's money supply is constant --
though continually-renewing via basic income distributions and demurrage --
should keep \coin ultimately ``anchored''
in its relation to time as discussed in \cref{sec:meta:time},
as fiat currencies gradually drift via inflation.

\subsubsection{Purchasing Power Analysis.}
\Cref{sec:purchasing} employs classical
inflation theory~\cite{feenstra2017international}
to analyze the purchasing power of \coin.
We find that \coin's purchasing power
may be expected to increase in the long run,
resulting in deflation,
whenever real economic growth exceeds population growth.
While deflation is deadly in classical economics,
this is because of the dependency of economies on debt-based money creation.
Monetary scarcity increases the real interest rates,
which disincentivizes borrowing and spending,
which makes money even more scarce.
We argue in \cref{sec:purchasing} that because
new \coin is created via debt-free basic income rather than loans,
deflationary spirals are unlikely to occur \coin
even in the presence of deflation.

Our exchange rate and purchasing power analyses are currently
based on the assumption that \coin has saturated a fairly large community.
We do not expect this assumption to hold
in a rapid-growth early phase development of \coin, however.
We speculate that as users adopt \coin
so that more and more goods and services can be purchased with \coin,
its exchange rate and purchasing power with respect to other currencies
would increase.
We have not yet modeled this scenario, however,
a task we leave to future work.

\subsubsection{Speculation and Saving Analysis.}
\Cref{sec:speculation} analyzes
how a rational individual would behave
with respect to saving or speculative HODLing of \coin
versus spending for productive use.
Due to the early adoption reward mechanism discussed above, 
speculation on \coin is likely to happen in early stages,
potentially making the currency unstable but also attracting more users.
However, after \coin has successfully saturated its potential user base, 
speculation is subject to tax via demurrage and therefore is disincentivized. 
Nevertheless, we find that the rich
might still save income to improve utility --
thereby paying higher tax rates than the poor --
even though a consistent global demurrage rate is applied to everyone.
Thus, we by no means expect speculation or savings to disappear
even once participation in \coin stabilizes.
To the extent it continues, however,
all participants are effectively compensated for any resulting instability
via the tax-and-redistribution effect of demurrage and basic income.

\cb{Big problem IMHO: After reading this section, I've got no idea which assumptions deviate from reality (as promised at the beginning of the sec) and how important these deviations are. Do other economic models make similar assumptions? Would my comment in the beginning of this section address it?}

\subsubsection{Stability versus Equitability.}
A central bank's primary mission is to maintain price stability,
traditionally by monitoring real economic indicators
and using a variety of policy levers to target
about 2\% inflation~\cite{federal15why,noyer16thoughts}.
Stablecoins~\cite{lipton2020tether,moin2020sok}
are cryptocurrencies that similarly pursue stability,
typically by pegging their value to that of a traditional currency --
and hence indirectly relying on that currency's underlying central bank.
All of these (direct and indirect) stabilization techniques depend on
complex economic monitoring and adjustment mechanisms,
none of which have yet proven stable \emph{in fact}
over historical periods~\cite{reinhart11this}.
\Coin follows Bitcoin's audacity of adopting an
``automatic monetary policy
based solely in nominal data''~\cite{bohme15bitcoin}.
This choice makes \coin's policy attractively simple,
while carrying the the immediate implication that -- like Bitcoin --
we cannot expect \coin to exhibit price stability
with respect to real economic activity
in the way that central banks and stablecoins aim to.

The dynamic controls that would be necessary for conventional price stability,
in fact, appear incompatible with \coin's mandate
of equitability in space and time,
at least as modeled above in \cref{principles}.
Adjusting the demurrage rate $\alpha$ to track economic indicators
would mean that the aggregate basic incomes distributed at some times
must represent a different proportion of total monetary wealth
than the aggregate basic incomes distributed at other times,
hence potentially eroding one generation's economic opportunity versus another.

Thus, the choice between price stability and equitability
as a currency's ``prime directive''
may represent a fundamental and in some sense irreconcilable difference.
Nevertheless, an intriguing question for future exploration
is whether a currency like \coin
might achieve a different form of stability in the long run:
\eg, if its user population grows sufficiently large,
if that population's demand for \coin (\eg, the average ``basket of goods'')
becomes sufficiently stable in a human behavioral sense,
and if demurrage limits inequality and disincentivizes speculation
sufficiently to ensure that \coin's long-run real value mostly reflects
relatively-stable aggregate human behavior of the population
and not the speculative sentiments of commercial banks and rich investors.
We leave this fascinating question 
of what ``stability'' really means
for future exploration.

\subsubsection{Migration Incentives.}

Finally, while so far only based on informal analysis,
we observe certain striking differences between
the borderless, permissionless basic income in \coin
and a conventional basic income proposal
implemented as a fiscal policy in some government jurisdiction.
Conventional basic income proposals
not only require making difficult choices about
what level of basic income is ``affordable''
balancing standard-of-living expectations against budget constraints,
but also can create resistance from the fact that
they incentivize to ``inward migration''
towards jurisdictions that have (larger) basic incomes,
potentially exacerbating
already-inflamed divisions and ``fortress'' mentalities.
The basic income embodied in \coin, in contrast,
promises only equitability rather than any particular standard of living --
but also ensures that its reward for participation
is borderless and available anywhere \coin can be adopted.
Further, because \coin's basic income will trade at the same value
against other currencies anywhere,
its \emph{purchasing power} will be greater in poorer regions
with lower cost-of-living.
If \coin creates any migration incentives at all, therefore,
they will be from richer to poorer jurisdictions,
thereby potentially addressing one significant perceptual roadblock
to UBI adoption.

\com{	A real future work section should be a concise summary
	of many opportunities for future work pointed out in the paper,
	while this one currently points out just one.
	So for now at least, just merging that one point
	into the relevant analysis section above..

\section{Future Work}
\label{sec:future}


\com{
So far, our exchange rate or purchasing power analyses are based on the assumption that \coin has saturated in a fairly large community or area. However, this is not a valid assumption at the beginning or early-phase development of \coin. We speculate that as more and more goods and services can be purchased with \coin, its exchange rate to other currencies and purchasing power would increase. However, we are missing an appropriate model to demonstrate it, which we leave as future work.
}


\com{
\subsubsection{UBI Comparison and Migration Incentive.} \coin essentially provides a possible bottom-up method of implementing the Universal Basic Income(UBI) on a global scale without relying on the traditional jurisdiction system. While a jurisdiction-based UBI would inevitably incentivize people to migrate to countries with higher UBI and living standard, a \coin-based UBI at a global scale would generate the opposite migration incentive: because everyone receives the same amount of nominal basic income, the basic income would have more purchasing power in areas or countries with cheaper goods and services. We leave the detailed model and analysis as future work.
}

}

\section{Related Work}
\label{sec:related}

\coin may be considered a digital
community currency~\cite{lietaer2006community,seyfang13growing},
intended as an experimental tool to support commerce and economic empowerment
among users who voluntarily opt into using it.
\com{
Most community currencies have focused on supporting
geographically localized communities,
and \coin might be most readily launched and used initially in this way too,
although there is nothing in its economic principles or design
that necessarily restricts it to localized use.
Different community currencies have been motivated by varying objectives,
supported by different mechanisms and issuance procedures.
}
Gesell's demurrage ideas~\cite{gesell58natural}
inspired the W\"ara and W\"orgl community currencies
during the Great Depression~\cite[chapter 13]{suhr89capitalistic} --
as well as the
WIR in Switzerland~\cite{lipton2020tether},
which still exists as a commercial barter network.
Community currencies in the W\"ara and W\"orgl tradition,
inter-enterprise currencies like WIR,
and cryptocurrencies like Bitcoin
all fit under the broad umbrella of
complementary currencies~\cite{meyer18money},
as does \coin as a new democratically-motivated
hybrid design point in this space.

Many have observed that time itself can serve as a currency.
Time banking was first introduced in Japan
by Teruko Mizushima~\cite{miller08teruko},
then followed by many time-based currencies globally.
While some time currencies
such as Time Dollars~\cite{cahn92time}
ask users to trade or bank one hour of anyone's time
at a fixed 1-to-1 ``exchange rate'',
other time currencies such as Ithaca HOURS
account for expertise by explicitly permitting specialists
to charge several time-currency HOURS per real hour
of their time~\cite{glover95hometown,hermann06special,jacob12social}.
\Coin builds on time as an ultimate foundation
for semantic meaning and value in a complementary currency,
while taking the pragmatism of Ithaca HOURS further by accepting
that the trade value of one \coin may freely ``float''
based not only on specialization but innumerable other adjustment factors
as discussed in \cref{sec:meta:time}.

Bitcoin inspired many new approaches for implementing
digital complementary currencies usable globally
via decentralized architectures.
Nimses~\cite{nimses17,larkina17nimses}
attempts to create a decentralized time-based cryptocurrency,
for example.
Encointer~\cite{brenzikofer2019encointer} and
GoodDollar~\cite{gooddollar} are recent cryptocurrencies
incorporating UBI principles.
Researchers have also analyzed various monetary aspects of cryptocurrencies,
such as the implications of
Bitcoin~\cite{bohme15bitcoin,lipton18blockchains},
Proof-of-Stake systems~\cite{viswanath2019compounding,brunjes2018reward},
stablecoins~\cite{lipton2020tether,moin2020sok},
and Central Bank Digital Currencies or
CBDCs~\cite{fung16central,lipton18blockchains,kumhof18central}.
To our knowledge, \coin is the first attempt at
designing and economically analyzing a permissionless cryptocurrency
motivated foremost around supporting
the democratic principle of equal opportunity,
overriding even the traditional primary goal of monetary price stability.

Other work has focused on the important challenge of
implementing \emph{Proof-of-Personhood}~\cite{borge2017proof,siddarth20who}:
\ie, a one-per-person notion of stake in a decentralized setting.
A key challenge here is addressing the false identity
or Sybil attack problem~\cite{douceur02sybil}.
\Coin builds on physical-world pseudonym parties~\cite{ford08nyms},
while other approaches build on
social trust networks~\cite{poupko20sybil},
biometrics~\cite{hajialikhani20uniquieid},
or government-issued identifiers~\cite{maram20candid}.
While security implementing decentralized proof-of-person\-hood
without compromising privacy
remains a critical unsolved problem,
it is orthogonal to and out of the scope of this paper.

\zhq{Related to the principle of fairness over time:
the study of intergenerational redistribution from
government debt.
\href{https://www.nber.org/papers/w3058.pdf}{The Politics of Intergenerational Redistribution}
\href{https://www.nber.org/papers/w21821.pdf}{THE POLITICAL ECONOMY OF GOVERNMENT DEBT}}

\section{Conclusion}

\coin is a cryptocurrency that aims to prototype
a more democratic, equitable, and sustainable form of money.
\coin introduces three principles to achieve its
goals: fixed basic income, equality over population,
and equality over total supply.
Through regular distribution of freshly-minted coins as basic income
to every participant,
\coin bounds inequality across the population.
Through a small demurrage rate of 2-5\% that slowly devalues each coin,
\coin controls the money supply and limits the lifetime of money,
bounding inequality across generations.
Using established economic models, our analyses on inequality,
exchange rate, purchasing power, speculation and saving show the
potential effectiveness and sustainability of \coin's monetary policy. 

\section*{Acknowledgments}

The authors wish to thank
Rainer B\"ohme, Xi Chen, Yawen Jeng, Luisa Lambertini,
Alexander Lipton, and Alexis Marchal 
for their extremely helpful comments and suggestions.
This research was supported in part by
U.S. Office of Naval Research grant N00014-19-1-2361
and by the AXA Research Fund.


\bibliography{main}
\bibliographystyle{splncs04}


\newpage

\appendix

\section{Discussion of Design Principles and Justifications}

This section discusses further issues and considerations
justifying \coin's design principles,
which were omitted from the main paper due to lack of space.


\subsection{Debt, economic growth addiction, and free money}
\label{sec:meta:growth}

It has been said that
``gold is the only financial asset that is not
someone else’s liability''~\cite{dalio20credit}.
In our current banking system based on fiat money,
both central and commercial banks can create money by issuing debt.
Fiat money thus always represents a liability to someone in the economy.
When a central bank expands its balance sheet
by purchasing assets such as government bonds or private debts,
the resulting money is the central bank's liability.
Commercial banks also create money ``out of thin air''
by issuing loans~\cite{werner2014can}.
When a commercial bank lends money to a borrower,
the bank creates assets and liabilities simultaneously,
thereby creating new money,
which is a liability of the commercial bank.

Fiat money therefore consists essentially of ``IOU''s:
any debt-based dollar existing in the economy implies
that someone will have to pay back this dollar,
generally with interest, sometime in the future.
Because the money needed to pay the compound interest on each loan
has not (yet) been created at the time the loan is issued, however,
we collectively face a ``grow or die'' problem:
the economy must perpetually grow in order for enough debt-based money
to exist in the future for all borrowers
to have any chance of paying off all the interest on loans issued now.
If the economy fails to grow rapidly enough
to cover the compound interest demanded on all existing loans,
then \emph{some} of those loans must default --
and in a major financial crisis
this often means \emph{many} loans
default~\cite{reinhart11this,helbing2019we}.

At its launch,
Bitcoin~\cite{nakamoto2019bitcoin}
appeared to create a remarkable new exception to the above rule.
``Anyone'' could mine new Bitcoin -- akin to mining gold --
without incurring debt or creating a liability for anyone.
It is unclear to what degree this exception to the money-is-debt rule
still holds,
now that mining is economically infeasible for anyone
without access to cheap power and the latest mining hardware,
and neither are readily available to newcomers \cb{Reads funny, perhaps the object is missing?  I do not understand what is not available to newcomers. The cheap power and the latest mining hardware? Then the sentence should be "and neither are *these* readily available...}
without incurring debt or another form of liability~\cite{vorick18state}.
Nevertheless, Bitcoin was merely the most recent demonstration
that it is \emph{possible} to create debt-free money
without incurring a liability on anyone,
as Gesell's ``free money'' theory had predicted~\cite{gesell58natural}
and as partly confirmed in subsequent
experiments and analysis~\cite{fischer33stamp,ilgmann11negative}.
Although both Gesell's stamp scrip and Bitcoin arose from other motivations,
breaking the unsustainable cycle
of economic ``growth addiction'' from debt-based money
may be one of the most important potentials they demonstrate in principle.

\subsection{Why regular distributions instead of one-time ``endowments''?}
\label{sec:meta:regular}

Why should distributions of money to support equal opportunity
be \emph{regular}, and not special ``one-time'' events for example?
As an alternative,
Condorcet~\cite{condorcet1794esquisse}
and Paine~\cite{paine1797agrarian} proposed in the 1790s
a one-time \emph{basic endowment} at birth or maturity.
Further, real governments have occasionally even implemented
one-time, ``more-or-less'' equal distributions
to ``more-or-less'' all citizens: \eg,
in the \emph{voucher privatization} programs
following the collapse of the USSR~\cite{boycko93voucher},
and more recently in emergency response
to COVID-19~\cite{karpman20unemployment}.

But steady-state economic reality is populated by a mix of people of all ages.
People are continually born, coming of age, and dying;
people are continually achieving wins and suffering failures and losses.
Real innovation requires risk-taking --
people need the opportunity to try something, fail, and start over.
One-time distributions can be lost through bad investments,
personal disasters such as addictions or other medical conditions,
displacement and losses from to war or natural disasters, etc.
In general, one-time distributions
cannot guarantee people, throughout their lives,
the power to start over with the same opportunities
they had before a loss.
This observation leads us to the conclusion that baseline support
must be a relatively \emph{continuous} -- or at least periodic --
supply made available to all individuals while still living
and potentially able to take advantage of it.

\subsection{On what basis to choose a reasonable demurrage rate?}
\label{sec:meta:demurrage}

An important question is how we might choose
the demurrage rate $\alpha$ in \coin.
We might approach this question on either a philosophical basis
or one of pragmatic historical experience.

We can find philosophical and moral foundations for demurrage
in the fact that many -- perhaps most --
forms of property ownership throughout history
have been time-limited rather than indefinite.
Most physical goods and capital have inherent time limits on ownership
by virtue of being perishable or wearing out:
food, materials, equipment, buildings, even land used unsustainably.
Counteracting money's durability advantage over other goods
was a primary motivator for Gesell~\cite{gesell58natural}.

For thousands of years, rulers of ancient Mesopotamian civilizations
including those of Sumeria, Assyria, and Babylon
regularly proclaimed acts of \emph{Misharum},
cancelling all debts and freeing debt-slaves empire-wide,
for economic renewal
and protection from encroaching aristocracy~\cite{hudson93lost}.
Jewish society encoded such a tradition into its most fundamental laws
by requiring a ``year of jubilee'' every 50 years.
Modern intellectual property law confers only limited-time ownership.
Shareholder corporations are fairly exceptional
in allowing unlimited-time ownership --
but there are strong arguments that firm ownership
should \emph{not} be unlimited,
as in proposals for stakeholder tenure~\cite{turnbull98should}
and other stakeholder governance models~\cite{letza04shareholding}.

Given this ancient and modern precedent alike,
it should not seem radical to view money as a social good
that society grants an individual possession of for a reasonable period,
but not forever.
In a cryptocurrency or smart contract,
we could certainly implement money that behaves like ``leprechaun gold''
by suddenly vanishing after a particular lifetime,
though doing so would compromise its fungibility
by making a coin's real value decrease with age~\cite{ford19money-anon}.
But observe that in a hypothetical steady-state ``leprechaun gold'' economy
a fraction $1/L$ of all money would vanish and have to be renewed year
if each coin lasts for $L$ years.
By eliminating $1/L$ of the value of \emph{all} coins each year via demurrage,
instead of eliminating \emph{all} of the value of $1/L$ of the coins each year,
we achieve the same aggregate rate of devaluation and renewal
while preserving the fungibility of coins.
Since $\alpha = 1/L$ is the demurrage rate,
we can consider the reciprocal of the demurrage rate
to be the effective \emph{lifetime} or \emph{ownership tenure}
of the demurraged coins,
even if they ``vanish'' only gradually rather than instantly.

But money is a social instrument used by \emph{people}
to reward \emph{people} for valued goods and services,
and the people so rewarded have limited lifespans
in which to enjoy (or further invest) those rewards.
There is then reasonable grounds
to tie the lifetime of such a reward to a time period
something like a human generation or working lifespan:
because that is the time in which rewards earned early in a person's life
may reasonably be expected to benefit \emph{them} --
rather than their heirs or successors --
whether through further investment or spending for enjoyment.
A reward that lasts significantly longer will primarily benefit
the rewardee's heirs or successors,
who will generally have different aptitudes and motivations,
and who may be unlikely to produce much (new) value for society
from that legacy reward.

From an equal opportunity perspective,
it is of course unjust for the children of poor families
to have their opportunities dominated and limited
by the economic losses or other hardships their ancestors faced.
But it is also arguably unjust to the children of rich families
to be denied ever knowing how much of whatever wealth they accumulate
truly reflects \emph{their own} accomplishments,
to be proud of,
and how much represents wealth and advantage they inherited
purely by luck of birth.
Thus, just as with a limited-duration intellectual property right,
tying the ownership tenure of demurraged money
to something comparable a working human lifetime
optimizes for conferring most rewards on \emph{those who earned it},
while allowing and expecting the descendents of both winners and losers
to prove themselves on a ``playing field'' that is, if not completely level,
at least not tilted to an unbounded and continually-growing degree either.

This is at least one philosophical basis for choosing a demurrage rate of, say,
somewhere between 2\% and 5\% per year,
corresponding to an ownership tenure period of
about a 50-year modern human working lifespan
or about a 20-year generation gap, respectively.
Gesell's proposed demurrage rate of 5.2\%,
though derived from the fact that there are 52 weeks per year,
comes out at the upper end of this range.
The ancient Mesopotamians, whose rulers most often
declared \emph{Misharum} on the ascension of each successive ruler --
i.e., about once per generation --
might be viewed similarly as precedent for the 5\% end of this range,
whereas the Jewish tradition's 50-year jubilee cycle
would be precedent for the 2\% choice.
(Granted, both human generations and lifespans
were substantially shorter then so the correspondence
is shaky and imprecise.)

The pragmatic historical basis reflects the experience
of modern central banks practicing Keynesian economics,
in which inflation-targeting serves a purpose closely-related to
if not quite identical to Gesell's demurrage~\cite{keynes65general}.
Experience seems to be that 2\% has proven
a reasonably safe and effective inflation target to steer between
the risk of deflationary spirals at the lower end,
and risk of overheating leading to hyperinflation or bank runs
at the upper end~\cite{federal15why,noyer16thoughts}.
There are also arguments for higher inflation targets
of around 4\%~\cite{ball14case}.
Thus, whether based on a fundamental basis of
matching ownership tenure to human lifetimes,
or based on the pragmatic economic experience of central banks,
demurrage rates in the 2--5\% range seem likely reasonable,
though there is nothing magical about any particular value.

\com{
Avoid the ``I want a pony!'' ``But ponies are expensive!'' debate.
}



\section{Long-Term Assumption and Adoption Incentive}
\label{sec:population}

\subsubsection{Long-Term Assumption on Population Changes.} 
We assume that in the long term, 
once \coin is widely deployed and adopted within some user population,
that population becomes relatively stable.
Mathematically, that is, $\exists \tau \in\mathbb{N}^+$ for $\forall t \geq \tau $,

\begin{equation}
\left| \frac{N_{t}}{N_{t-1}} - 1 \right| \leq \epsilon,\quad \epsilon \approx 0\,.
\end{equation}

In reality, participation in \coin will not be fixed or perhaps even stable,
especially in early phases
when we would hope to see rapid adoption.
In the long term, however, once \coin has saturated
whatever population proves amenable to adopting it,
we expect participation to stabilize since
human population changes slowly compared with most economic effects.

\subsubsection{Adoption Incentive and Long Term Supply Stability.}
\label{adoptionIncentive}


The dynamic interest rate mechanism creates a potentially strong 
adoption incentive in early stages,
as our previous discussions in \cref{sec:meta} showed that 
the interest rate $R_{t}=(1+n_t)(1-\alpha)-1$.
We illustrate this adoption incentive through an example. 
Suppose that the number of participants doubles in one period ($n_t=2$),
for example.
The interest rate $R_t$ would be close to 2, if $\alpha$ is sufficiently low,
implying that the nominal savings of ``early adopters''
nearly double in just one period.
This does not mean that \emph{real} value
necessarily doubles correspondingly, of course.
But if the new adopters put the currency
in active use and circulation similarly to the existing users,
thereby growing the real \coin economy roughly proportionally as well,
then the early-adoption reward will also be meaningful in real value.

When \coin successful saturates among all the potential users, with our long-term assumption on population size, the interest rate would drop to nearly zero,
\ie, the early-adoption reward tapers off and disappears.
Mathematically, a negative interest rates is equivalent to $\epsilon <\alpha/(1-\alpha)$ when $t\geq \tau$.

The long-term stability of \coin supply is
a significant distinction from the fiat currencies' supply,
which grows exponentially over time.

\section{Inequality Analysis}
\label{sec:inequality}

We utilize three different metrics to analyze how the monetary policy of \coin can have effects on inequality among all participants\footnote{For all participants together, we set $X_{t} =\{x^i_{t}\}$, for $i=1, 2, \dots, N_t$}. These metrics complement each other, in that some capture aspects overlooked by others. 
\zhq{We have not mentioned Gini is one of our metrics yet:For example, a known drawback of the Gini coefficient is that it overlooks the distribution of wealth (the same Gini value corresponds to many wealth distributions), but the inequality ratio metric captures the distribution.}\cb{@Haoqian: Added the 2 previous sentences, are they correct?} We emphasize that, it is one of the most significant distinct property from current fiat currencies that the design of \coin itself reduces the inequality and brings a theoretical upper bound of inequality.

\subsubsection{Reducing Inequality through Demurrage.} When the number of participants is stable and fixed, after each devaluation of the existing \coins and distribution of basic income to every participant, the inequality level reduces. We first measure inequality by both variance and the Gini coefficient. For the variance, the inequality is strictly reduced, since

\begin{equation}
\label{Variance}
    Var((1-\alpha)X+B)=(1-\alpha)^2Var(X)<Var(X) .
\end{equation}

In economics, the Gini coefficient aims to measure the inequality level within a group of people. The Gini coefficient $G$ is defined as half of the relative mean absolute difference of variables, \ie 

\begin{equation}
    G(X) = \frac{\displaystyle{\sum_{i=1}^N \sum_{j=1}^N \left| x_i - x_j \right|}}{\displaystyle{2 N^2 \bar{x}}} .
\end{equation}

We know from \cref{participants} that, when the number of participants is fixed, the \coin supply is fixed as well. Therefore, after the devaluation and distribution of the new \coin, the mean of the participants' balances remains unchanged. Hence, we obtain

\begin{equation}
\label{GiniCoefficient}
    G((1-\alpha)X+B)=(1-\alpha)G(X)<G(X) .
\end{equation}

Therefore, when the population size is stable, the \coin monetary policy reduces inequality measured in both variance and the Gini coefficient.

Next, we analyze the inequality level between any two participants. We use the ratio of their \coin balances to measure the inequality level:

\begin{definition}
For any pair $(i,j)$, $e_{i,j}$ is the inequality ratio between them, and
\begin{equation}
e_{i,j}=
\begin{cases}
\frac{x^i}{x^j}& \text{$x^i>=x^j$}\\
\frac{x^j}{x^i}& \text{$x^i<x^j$}\\
\end{cases}
\end{equation}

\end{definition}

According to the definition, when $e_{i,j}=1$, 
the two participants have the same amount of \coin balance, 
and a bigger $e_{i,j}$ means a higher inequality between the two participants. 
Without loss of generality, we assume $x^i>=x^j$ (otherwise we can switch $i$ and $j$). 
For any pair $(i,j)$, 
immediately after the demurrage and new basic income distribution,
we have

\begin{equation*}
    \frac{(1-\alpha)x^i+B}{(1-\alpha)x^j+B}\leq \frac{x^i}{x^j} = e_{i,j} .
\end{equation*}

Therefore, the inequality ration will not increase after the event, and if $x^i>x^j$,
the inequality is strictly reduced.

\subsubsection{Upper Bound on Inequality.}
We analyze the upper bound on inequality in terms of the
variance across all participants' balances, the Gini coefficient, and ratio between any two balances. 
To have a fixed sample size, 
we continue to assume the number of participants is stable and fixed. 
Because we investigate the worst situation, 
$X$ can be arbitrary 
with condition $x^i>=0$ for any i and $\sum_{i=1}^{n} x^i=M=\frac{1}{\alpha} B N$.

For any distribution of $X$ satisfying the conditions, the upper bound of variance is

\begin{equation}
    \sup_{X} {\rm Var}((1-\alpha)X+B) = (\frac{1-\alpha}{\alpha}B)^2(N-1) .
\end{equation}
\textit{(Proof in \cref{sec:inequality:variance})}
\newline

A Gini coefficient of one indicates maximal inequality. 
For large and finite size of participants, 
where only one person has all the \coins, 
and all others have none, the Gini coefficient will be nearly one. 
When the number of participants tends to be infinite, 
the Gini coefficient will be one. 
In our case, after the demurrage and distribution event, 
the Gini coefficient will be very close to $1-\alpha$. 
Specifically, we have

\begin{equation}
    \lim_{N \to \infty}\sup_{X} {\rm G}((1-\alpha)X+B) = 1-\alpha .
\end{equation}
\textit{(Proof in \cref{sec:inequality:gini})}
\newline

For the inequality ratio between any two participants, after a devaluation and basic income distribution, we have

\begin{equation*}
    \sup_{1\leq i,j \leq N} e'_{i,j} = \sup_{1\leq i,j \leq N} \frac{(1-\alpha)x^i+B}{(1-\alpha)x^j+B} = \frac{1-\alpha}{\alpha} N + 1 .
\end{equation*}
\textit{(Proof in \cref{sec:inequality:ratio})}

\subsection{Proof of Variance Upper Bound}
\label{sec:inequality:variance}

The optimization problem can be written as
\begin{eqnarray}
\label{ProofVariance}
    \text{maximize}_{X} && Var((1-\alpha)X+B) \\ 
	\text{s.t.}&& x^i>=0 \;, \forall i \in [N] \nonumber \\
	&& \sum_{i=1}^{N} x^i=\frac{1}{\alpha} B N, \nonumber
\end{eqnarray}
	
By applying properties of variance, we have

\begin{equation*}
    Var((1-\alpha)X+B) = (1-\alpha)^2Var(X) 
\end{equation*}

\begin{equation*}
    Var(X) = \frac{1}{N}\sum_{i=1}^{N} (x^i)^2 - (\frac{1}{N}\sum_{i=1}^{N} x^i)^2 = \frac{1}{N}\sum_{i=1}^{N} (x^i)^2 - (\frac{1}{\alpha} B)^2
\end{equation*}

Therefore, the optimization problem is equivalent to

\begin{eqnarray}
	\text{maximize}_{X} && \sum_{i=1}^{N} (x^i)^2 \nonumber\\ 
	\text{s.t.}&& x^i>=0 \;, \forall i \in [N] \nonumber \\
		&& \sum_{i=1}^{N} x^i=\frac{1}{\alpha} BN, \nonumber
\end{eqnarray}

Since $\sum_{i=1}^{N} (x^i)^2 \leq (\sum_{i=1}^{N} x^i)^2 = (\frac{1}{\alpha} BN)^2$ and the equality is achieved when $x^i=\frac{1}{\alpha} BN, i \in N$ and $x^j=0$, for all $j \in N$ and $j \neq i$. Substituting the results, we obtain the upper bound is $(\frac{1-\alpha}{\alpha}B)^2(N-1)$.

\subsection{Proof of Gini Coefficient Upper Bound}
\label{sec:inequality:gini}

We first prove the upper bound with fixed $N$. The optimization problem can be written as 

\begin{eqnarray}
\label{ProofGini}
    \text{maximize}_{X} && \frac{\displaystyle{(1-\alpha)\sum_{i=1}^N \sum_{j=1}^N \left| x^i - x^j \right|}}{\displaystyle{2 N^2 {x}}} \\ 
	\text{s.t.}&& x^i>=0 \;, \forall i \in [N] \nonumber \\
	&& \sum_{i=1}^{N} x^i=\frac{1}{\alpha} BN. \nonumber
\end{eqnarray}

The denominator is constant with a fixed $N$ and $\sum_{i=1}^{N} x^i=\frac{1}{\alpha} BN$. Let us assume that $x^1 \geq x^2 \geq ... \geq x^{N-1} \geq x^N$, we have

\begin{equation*}
    \sum_{i=1}^N \sum_{j=1}^N \left| x^i - x^j \right| \leq 2\sum_{i=1}^N \sum_{j=i+1}^N x^i \leq 2(N-1) \sum_{i=1}^N  x^i = \frac{2(N-1)}{\alpha} BN.
\end{equation*}



The equality is achieved when $x^1=\frac{1}{\alpha} BN$ and $x^i=0, 2 \leq i \leq N$. Substituting the results, we obtain the upper bound is $(1-\alpha)\frac{N-1}{N}$. When $N$ tends to infinity

\begin{equation*}
    \lim_{N \to \infty}(1-\alpha)\frac{N-1}{N} = 1-\alpha .
\end{equation*}

\subsection{Proof of Ratio Upper Bound}
\label{sec:inequality:ratio}

For $N \geq 2$, let us assume that $x^1 \geq x^2 \geq ... \geq x^{N-1} \geq x^N$, we have

\begin{eqnarray*}
    \sup_{X, 1\leq i,j \leq N} \frac{(1-\alpha)x^i+B}{(1-\alpha)x^j+B} &=&
    \sup_{X} \frac{(1-\alpha)x^1+B}{(1-\alpha)x^N+B} \\
    &\leq& \frac{(1-\alpha)(\frac{1}{\alpha} BN)+B}B \\
	&=& \frac{1-\alpha}{\alpha} N + 1\;.
\end{eqnarray*}

The equality is achieved when $x^1=\frac{1}{\alpha} BN$ and $x^i=0, 2 \leq i \leq N$.

\section{Exchange Rate Analysis}
\label{sec:exchange}

\subsection{Long-run Analysis}
\label{exchange}

First, we focus on formulating the long-term exchange rate of \coin
to other fiat currencies.
For this purpose,
we assume that
(a) \coin has successfully saturated a sufficiently large community
for general transactions,
(b) price is flexible in the long run, and
(c) that \textit{Purchasing Power Parity} (PPP) holds in the long run.
We adopt the PPP theory \cite{a19901}
to model \coin's long-term exchange rate.\footnote{
	Empirical evidence \cite{feenstra2017international} indicates that
	the exchange rates of several currencies against U.S. dollars
	have reached their equilibrium,
	exactly as the PPP theory has predicted.
}
The PPP theory argues that two currencies are in
equilibrium when a basket of goods is priced the same in both currency area,
taking into account the exchange rates. Mathematically, the exchange rate can
be formulated as a function of several important economic indicators, \ie,
\begin{equation}
\label{equilibriumExchangeRate}
    E_{\popsign/\foresign}=\frac{P_{\popsign}}{P_{{\foresign}}}=\frac{M_{\popsign}/M_{{\foresign}}}{{\bar{L}_{\popsign}Y_{\popsign}}/{\bar{L}_{{\foresign}}Y_{{\foresign}}}}\,
    ,
\end{equation}
where $M$, $P$, $Y$ and $L$ denote the currency quantity, the price level, the real income and the liquidity demand of the currency, respectively.
The subscript $\popsign$ denotes \coin
and $\foresign$ denotes the compared fiat currency.
$E_{{\popsign}/{\foresign}}$ is the exchange rate between the two currencies.
Here we further assume a fixed liquidity demand in the long run analysis
and re-denote it by $\bar{L}$.



Money supply
for a fiat currency does not have a single ``correct" measure.
Instead, it has several definitions for various purposes,
for different countries or under their accounting rules.
This section uses M2 as the supply of fiat currency,
including cash, demand deposits, and saving deposits in banks.
M2 is usually the key economic indicator
forecasting inflation \cite{hallman1991price}.

By contrast,
cryptocurrency generally has a clear definition of money supply.
Anyone can easily compute the total supply of Bitcoin at any time,
for example, based on the block number.
In \coin, anyone can trivially  calculate the total \coin supply
based on \cref{participants}.
The ongoing development of decentralized finance (DeFi)
will likely lead to other various definitions of money supply
in the future.
In any case, cryptocurrency and fiat money have fundamental differences: the cryptocurrency deposit in banks will never be like the cryptocurrency recorded in the decentralized blockchain since there is no central bank or government to bail out the banks. 
\baf{ How is "no central bank to bail out the banks" a relevant difference here?
	Or isn't it more the fact that cryptocurrencies represent
	"full-reserve" currencies that is more relevant here?}

The trend of money supply for fiat currencies
has always been increasing so far.
The plot of the USD M2 money stock
provided by Federal Reserve Economic Data (FRED), for example,
clearly shows this trend.\footnote{\url{https://fred.stlouisfed.org/series/M2}}
On the other hand,
cryptocurrency generally has a clear policy for its supply,
maintained by machines enforcing software-encoded rules.
It is therefore easy to forecast the future supply of cryptocurrency.
This is an important feature as it reduces complexity
in predicting the exchange rate of crytocurrencies relative to a fiat currency.

The \coin exchange rate relative to fiat currencies
is formulated as below by transforming \cref{equilibriumExchangeRate}
into its relative form

\begin{equation}\label{eqn.relativeform}
    d_{{\popsign}/{\foresign},t}=(\mu_{{\popsign},t}-\mu_{{\foresign},t})-(g_{{\popsign},t}-g_{{\foresign},t})\,,
\end{equation}
where $\mu$ is the growth rate of currency supply,
and $g$ is the real income growth rate.
This equation formulates the change of the exchange rate $d$
as the difference of currency growth and real income growth
in both currency areas.
If we assume that both currency areas have the same real income growth,
the exchange rate is completely determined by the difference of currency growth.

Recall that \cref{participants} states that
the only factor dominating the supply of \coin, other than time,
is the number of participants or population size.
Thus, the growth rate of \coin is the same as population growth rate.
If we further assume that the change of population is limited
and relatively slow,
the supply of \coin will be similarly constant and stable.
\com{	seems redundant?
In the case of a massive population change of the population,
the supply of \coin will be expanded or contracted proportionally.
In other words, \coin's growth rate is the same as population growth rate.
}
With the assumption that growth rate is stable in the long run,
\cref{eqn.relativeform} takes this reduced form:

\begin{equation*}
    n_{{\popsign},t}-\mu_{{\foresign},t}=d_{{\popsign}/{\foresign},t}
\end{equation*}

The end result is simple: as long as the population size is stable
and the fiat currency supply expands as usual,
the long-term level of the exchange rate of \coin
relative to inflationary fiat money will always increase over time.

\subsection{Short-run Analysis}
\label{short-run}

\cb{Perhaps for a next iteration: it seems that, in the short run, the interest rate of two currencies closely matches their exchange rate. Might be worth to explore.}

In this subsection, we briefly demonstrate a model for short-run analysis.
We adopt an asset approach to explain the exchange rates
in the short run \cite{feenstra2017international}.
In this approach, we consider all currencies to be assets,
and therefore currency holders
could have capital gains on their currencies
by participating in a lending market.
For \coin, its nominal interest rate depends on two factors:
(a) the interest rate from the \coin mechanism and
(b) the interest rate from lending.
At the end of this subsection,
the short-run analysis gives us a guideline
on how to reduce fluctuation of the short-run exchange rate.
\baf{ what is this last sentence trying to say, and why is it relevant here?}

The short-run analysis relies on the long-run analysis and its assumptions.
We further assume that
(a) the exchange rate predicted by the long-run analysis
is the expected future exchange rate,
(b) the price level in the short run is sticky,
(c) there is a sound lending market, and
(d) the nominal interest rate is flexible
and \textit{Uncovered Interest Parity} (UIP) holds in the short run.

The UIP theory states that the exchange market is in equilibrium
when the expected rates of return on each type of currency investment are equal.
Mathematically, this can be formulated as

\begin{equation}
\label{UIP}
    i_{\popsign} = i_{{\foresign}} + \frac{E^e_{{\popsign}/{\foresign}}-E_{{\popsign}/{\foresign}}}{E_{{\popsign}/{\foresign}}}\,,
\end{equation}

where $i$ denotes the nominal interest rate of the currency, and the superscript $e$ means the expected value in the future.
\baf{ Would it be better to use $\mathbf{E}[x]$ for the expectation of $x$,
	as I think is more traditional notation for expectation?
	I know it's a potentially-confusing second use of the letter E,
	but that can be mitigated by putting it in a different font.
	But if the superscript-e notation is common in economics,
	then that's fine too.}
Based on our assumption,
the expected exchange rate is from our long-run analysis or 

\begin{equation}
\label{equilibriumExpectedExchangeRate}
    E^e_{{\popsign}/{\foresign}}=\frac{P^e_{\popsign}}{P^e_{{\foresign}}}=\frac{M^e_{\popsign}/M^e_{{\foresign}}}{{\bar{L}^e_{\popsign}Y^e_{\popsign}}/{\bar{L}^e_{{\foresign}}Y^e_{{\foresign}}}} .
\end{equation}

In contrast to the long-run analysis,
in which we assume that the price level is flexible
and prices adjust to bring the market to equilibrium,
in the short run we assume price is sticky,
and it is the adjustment of nominal interest rates in each currency zone
that brings the money supply and money demand into equilibrium.
Hence, unlike in the long run,
we assume that $L$ is a decreasing function
of the nominal interest rate of $i$.
Mathematically, for a currency, we have

\begin{equation}
\label{shortrunPrice}
    \frac{M}{\bar{P}}=L(i)Y .
\end{equation}

We now have all the building blocks
to predict the exchange rate of \coin in the short run.
When \coin has successfully saturated in a community
with a stable population size,
implying its supply is stable and predictable,
with all else equal,
a permanent issuance of the compared currency
would influence the exchange rate in both the long run and short run.
In the long run,
we know that an increase of $M^e_c$
caused by the increase of the currency $c$
leads to the expected exchange rate $E^e_{{\popsign}/{\foresign}}$ decreasing.
In the short run,
as the price level is fixed,
according to the \cref{shortrunPrice},
its nominal interest rate will decrease.
Both effects would lower the spot exchange rate
according to the \cref{UIP}.

Denoting the new spot exchange rate as $E'_{{\popsign}/{\foresign}}$,
it is not hard to prove that
$E'_{{\popsign}/{\foresign}}<E^e_{{\popsign}/{\foresign}}$,
a phenomenon that economists refer to as \textit{exchange rate overshooting}.
The analysis tells us that when there is a tendency
for more permanent monetary policy shocks,
then there will be a tendency
for a more volatile exchange rate~\cite{feenstra2017international}.
This suggests that having a fixed monetary policy as in \coin
could potentially help to reduce volatility of the exchange rate.

\section{Purchasing Power Analysis}
\label{sec:purchasing}

We analyze \coin's purchasing power by formulating its inflation rate.
Inflation reduces a currency's purchasing power
as the prices of goods and services increase.
This is a long-run analysis
based on the assumption made in \cref{exchange},
and is not necessarily applicable in the short run.

We begin our analysis with
the quantity theory inflation equation~\cite{feenstra2017international}, \ie,

\begin{equation*}
    \pi_{t} = \mu_{t} - g_{t}\,,
\end{equation*}

where $\pi$ is the inflation rate,
$\mu$ is the growth rate of the currency supply, 
and $g$ is the real income growth rate in its currency zone.
As we stated earlier,
\coin's nominal growth rate is equal to the growth rate of the population
in its currency area. Therefore,
we may apply the equation to \coin as

\begin{equation*}
    \pi_{t} = n_{t} - g_{t}\,,
\end{equation*}

where $n$ is the population growth rate,
$n_t = N_t/N_{t-1} - 1$. 
This equation simply tells us that,
if the population is relatively stable in the long run
($n_t$ is 0 or very close to 0)
and the real growth rate is positive,
\coin will deflate rather than inflate.
With little or no real income growth,
the price level denominated in \coin will be stable.

Thus, in contrast with the mainstream tradition
of central banks targeting
a mild inflation rate~\cite{federal15why,noyer16thoughts},
\coin is deflationary in the long run
whenever the economy's real growth rate
exceeds population growth rate. 
Is this a bug or a feature?
In classical economics it would definitely be a bug,
due to the risk of deflationary spirals.
But would the same risk apply to a \coin economy?

\subsubsection{The risk of deflationary spirals.}
We argue that a classic deflationary spiral is unlikely to occur in \coin,
and hence that deflation is not necessarily bad for \coin,
because of the way it creates money via basic income rather than debt.
This conclusion may be true as well
for other cryptocurrencies whose issuance does not rely on debt,
such as Bitcoin's early period
before mining became power- and capital-intensive~\cite{vorick18state}.

To understand why, we have to examine how today's fiat money
is ``printed". As we stated earlier in this section, there are many different
measures of the currency supply. Why there are so many? One of the reasons is
that not only central banks can ``print" money,  but also commercial banks can
``print" money \cite{werner2014can,mcleay2014money}. More specifically,
monetary base (MB) is created or directly controlled by the central bank, while
M2 includes the money created by commercial banks.

How is money created and distributed? In cryptocurrency, this creation is simply
written in its code. For instance, Bitcoin is created whenever a new block is
mined, and the newly issued bitcoins are distributed to the miner of the block,
while \coins are created regularly and distributed to every participant
equally. In the fiat money system, no one solves mathematical puzzles, nor
does everyone receive an equal portion of newly created currency.
How does it work in fiat currency?
A simple and infeasible solution is to give the newly created
money to the government directly. However, this power is so easy to be abused
and nowadays the majority of countries have more or less independent central
banks making sure that newly created currency will not directly put into the
government's account. 

The real ``secret" is that today's fiat money is based on credit. Money is
created whenever an entity borrows money from commercial banks or central
banks. This entity could be a government that wants to fund its deficit, an
individual who takes a mortgage, or even a bank (including the central bank)
itself which creates money from the thin air by borrowing money to itself to
fund its purchase. Meanwhile, when the borrower repays, the principal of loans
is destroyed~\cite{lipton2018blockchains}. This implies that the key distinction
from cryptocurrency is that cryptocurrency lending does not create new money in
general.

How does the mechanism of fiat money creation relate to deflation?
Here we adopt an analysis by Fisher in 1911~\cite{assous2013irving} to
illustrate the deflationary spiral and to argue why \coin and cryptocurrency do
not suffer from it. We begin the spiral by a fall of the price level or
deflation. This causes the rise of the real interest rate, increasing the cost
of borrowing.
This urges borrowers to reduce their demand for loans,
causing the money supply to fall.
All else being equal,
reduced currency supply means price level will decrease even further,
and therefore a deflationary spiral has formed.

In contrast to fiat money, however,
\coin is not based on credit.
A fall in the price level does not influence the \coin supply,
because new money is injected constantly via debt-free basic income
rather than through loans.
Although deflationary spirals are a major risk for fiat currencies,
therefore,
we have substantial reason to believe that \coin
and other non-debt-based currencies
could substantially mitigate this risk.

\section{Speculation and Saving Analysis}
\label{sec:speculation}

As with other cryptocurrencies,
speculative investment is likely to happen in \coin.
This may cause unpredictable exchange rate swings in the short run,
as in today's market for unpegged cryptocurrencies.
Is this a bug or a feature?

We reiterate that in \coin,
price stability is not the primary goal but only a secondary goal
subsidiary to long-term fairness and equity,
as discussed in \cref{sec:meta:equality}.
Thus, while stability would be nice to have if and when we can get it,
we are willing to live with some instability --
especially short-term instability --
if doing so helps us reach the currency's long-term monetary policy objectives.

To analyze susceptibility to speculation,
it is useful to distinguish between three situations that might prevail
with respect to the participating user population:
rapid growth, stability, or rapid degrowth.

\com{ redundant with below?
In the early phase, speculative investment
is encouraged by the early-adoption reward mechanism, while after \coins
saturated among the population, speculative investment is discouraged by the
demurrage mechanism. Nevertheless, it is still rational for a participant to
save his \coin for the future, despite that a demurrage fee is applied. We
prove that rational rich and poor people would adopt different saving
strategies \ie~the rich pay more tax than the poor, even though the global
interest rate is the same for all.
}

\subsubsection{Early-adoption rewards dominant during rapid growth.}
First consider the situation in which
participation in \coin is growing rapidly,
as might occur in an early phase if it achieves
a critical mass of interest to drive rapid adoption
as happened to Bitcoin in its early years.
In such a phase,
speculative investment is encouraged
by \coin's ``early-adopter's'' reward mechanism
as described in in \cref{adoptionIncentive}.
With belief that more participants will adopt \coin scheme in the future,
current participants can expect a positive reward
and gain more \coin from HODLing than spending it.
With this adoption incentive, people may be more willing to adopt \coin,
and to convince others to adopt it as well,
which may be exactly what is desired in such an early phase.
Because the early-adopter's reward tapers off to nothing
as participation growth slows or eventually stops,
however,
this reward mechanism --
and the speculative swings and bubbles it might contribute to --
should be temporary and self-limiting.

\subsubsection{HODLing tax during periods of population stability.}
Once an early rapid-growth phase stabilizes
and \coin has saturated whatever community is amenable to adopting it,
we expect \coin's demurrage to take over eventually
and disincentivize too much speculation by effectively ``taxing'' it.

We utilize the model and assumptions in \cref{short-run} to illustrate the
influence of speculative investment in the short run on its exchange rate to
other currencies.
We model the HODLing consequence as a temporary shock of
reducing the \coin supply, assuming all other exogenous variables remain
unchanged and fixed. As \cref{shortrunPrice} showed,
the nominal interest rate would increase caused by the temporary shock to \coin.
As the expected future \coin supply remains fixed,
\cref{UIP} shows that the spot exchange
rate of $E_{{\popsign}/{\foresign}}$ decreases,
leading to the appreciation of \coin.

While the appreciation of \coin benefits the \coin community members who
purchase goods and services denominated in other currencies,
speculators need to pay the tax -- \ie, the demurrage fee --
under the assumption that the population size of the \coin community
is reasonably stable.
Therefore, speculating on \coin is discouraged
after a rapid-growth phase stabilizes.

However, if we consider the utility of a rational participant,
we obtain a different result:
A rational participant might be willing to pay a demurrage fee
for keeping his \coin to maximize his utility as a whole.

\subsubsection{Rational Saving and Proportional Tax.}
A rational participant
with prefect foresight may save his \coin for the future,
even though \coins are subject to lose their value by demurrage mechanism.
We prove that the \coin mechanism can automatically distinguish
rich and poor and apply different proportional tax rates of their saving,
even though everyone is subject to the same global interest rate.

A rational participant may save some of his income
by reducing the consumption today to increase the consumption in the future,
even when the saving is subject to lose some of its value by demurrage.
Let us consider a simple world,
where agents with prefect foresight who can only live for two periods,
of which he is young in the first period and old in the second period.
The agents are able to have earned
income apart from basic income $B$ when young,
and only basic income when old.
We use $in^i_t$ to denote the earned income of the agent $i$
at period $1$ and $out^i_t$ to represent his expenditure
at period of $t$.
Therefore, at the end of period 1, his balance is

\begin{equation}
\label{period1}
    x^i_{1} = B+in^i_{1}-out^i_{1} .
\end{equation}

His balance in period is subject to change by a factor of $R_2$ (the global interest rate at period $2$) due to the mechanisms of demurrage and early adoption reward. Hence, at the end of period $2$, his balance is

\begin{equation}
\label{period2}
    x^i_{2} = x^i_{1}(1+R_2)+B-out^i_{2} .
\end{equation}

Because the agent knows that he can only live for two periods, it is rational for him to spend all his \coin before the end of period two. Therefore, we have

\begin{equation}
\label{period2balance0}
    x^i_{2}=0 .
\end{equation}

By combining \cref{period1}, \cref{period2} and \cref{period2balance0}, we can derive his budget constraint as

\begin{equation}
\label{spending}
    out^i_{2} = (B+in^i_{1}-out^i_{1})(1+R_2)+B .
\end{equation}

The problem he is facing is to maximize utility of consumption in real term. Suppose that the utility is give by $(\frac{out^i_{1}}{P_{1}})^{1/2}+(\frac{out^i_{2}}{P_{2}})^{1/2}$, we can find the utility as the following function of $out^i_{1}$:

\begin{equation}
\label{utility}
    (\frac{out^i_{1}}{P_{1}})^{\frac{1}{2}}+(\frac{(B+in^i_{1}-out^i_{1})(1+R_2)+B}{P_{2}})^{\frac{1}{2}} .
\end{equation}

We can find the maximum value by differentiating this function with respect to $out^i_{1}$, and set the derivative equal to
zero. The utility function achieves maximum value when

\begin{equation}
\label{maxium}
    out^i_{1} = \frac{(1+R_2)(B+in^i_1)+B}{(1+R_2)^2\frac{P_{1}}{P_{2}}+(1+R_2)}
\end{equation}

In order to better understand this formula, let us consider some special cases. For simplicity, we assume that ${P_{1}}={P_{2}}$. For an agent with zero earned income, his consumption is equal to $B/(1+R_2)$ in the first period. When $R_2 = 0$, his consumption is equal to his basic income \ie he will consume all of his income in the first period. If $R_2 > 0$, he can improve his utility by consuming less than his basic income in the first period, so that he can receive the extra reward in the second period. If $R_2 < 0$, the best option for him is to consume more than his income. However, if borrowing is not possible for him, the best decision is to consume all his basic income. 

Now, let us consider the opposite situation. This time, we assume a agent belonging to the high-income group, in which they are able to have super high earned income in the first period \ie \ $in^i_1 \gg B$. From \cref{maxium}, his first period consumption is approximate equal to $in^i/(2+R_2)$. When $R_2 = 0$, he is about to save half of his income, and when $R_2 > 0$, he is going to save even more to enjoy the reward. If $R_2 < 0$, although his saving is decreased, it is still rational for him to save a significant amount of \coin,
so that he can improve his consumption in the second period,
even if the savings are subject to lose some of their value.

Therefore, if we consider utility,
a participant may rationally save some of his income
so that he can increase his utility as a whole.
Hence, the rich pay higher tax rates than the poor,
even they are subject to the same global demurrage rate.
\Coin therefore automatically distinguishes the rich and poor
and charges different tax rates according to their income level.



\subsubsection{The risk of rapid degrowth.}
We may worry that there is a risk that
if participation in \coin ever starts decreasing,
this could lead to a different form of positive feedback loop
or ``death spiral''
due to the effect opposite that of the early adopter's reward above.
That is, if participation is decreasing,
a basic income acquired at time $t-1$
will tend to be worth less if it is saved until time $t$.
Would this trigger more participants to leave the system,
increasing the rate of participation degrowth, and so on?
While this is an issue worth considering,
it seems unlikely to be a major problem for one simple reason:
basic income is \emph{free} to all participants.
So even if participation drops for whatever reason
and the remaining participants suffer an ``early adopter's penalty'',
all those remaining participants, if rational,
still gain more by continuing to participate
than by dropping out and giving up their basic income.
For this reason, the main risk we perceive
to \coin collapsing is not through a participation death spiral
but through other reasons,
such as because interest and economic activity using \coin becomes too weak
and people stop using it simply because
it is not useful or valuable enough to be worth the effort to participate.

\section{\coinlets: an Internal Microcurrency to Avoid Wallet Redenomination}
\label{sec:poplets}

\begin{algorithm}[t!]
\caption{Pseudocode implementation of \coin with \coinlet}
\label{alg:poplet}

\SetCommentSty{textnormal}
\DontPrintSemicolon
\SetKwInOut{Input}{input}
\SetKwFor{For}{for}{}{end}

\Input{$B$, the number of \coins issued to each participant per minting}
\Input{$\alpha$, the fraction of total \coin supply redistributed per minting}
\BlankLine
$N_0 \leftarrow$ initial number of participants at launch\;
$E \leftarrow$ a constant
    \tcp*{initialize the exchange rate}
\For(\tcp*[f]{one minting per time period}){$t = 1$ \KwTo $\infty$}{
	$N_t \leftarrow$ number of participants at time $t$
		\tcp*{take new participation census}
	$E \leftarrow E (1-\alpha) N_t / N_{t-1}$
		\tcp*{update the exchange rate}
	issue each participant $B/E$ \coinlets
		\tcp*{distribute new basic income}
}
\end{algorithm}

\Coin's definition and monetary policy as defined by \cref{alg:popcoin}
has the technical drawback that \emph{all} wallet balances on the ledger --
not just the one wallet per participant that receives new basic income --
must be adjusted for participation changes and demurrage in each period.

For implementation convenience and efficiency,
we may prefer an alternative method
of enforcing \coin's monetary policy
without affecting the nominal values of all wallet balances at each minting.
We can achieve this goal by introducing
a closely-related currency we will call \coinlet,
which has a time-varying exchange rate with \coin.
The exchange rate is maintained by software,
so \coin's users do not need to be aware of the existence of \coinlets.

In Bitcoin, the \textit{Satoshi} represents
the smallest atomic unit of Bitcoin that can be transferred.
Similar to the Satoshi,
the \coinlet is the indivisible atomic unit in \coin.
Unlike Bitcoin,
whose``exchange rate'' between Satoshi and Bitcoin is constant,
we adopt a time-varying exchange rate between \coinlet and \coin,
so that we do not need to update wallet balances at mintings.
The re-basing mechanism is achieved by changing the exchange rate
between \coinlet and \coin, as demonstrated in \cref{alg:poplet}. 
In short, \coinlet is an inflationary but participation-independent currency,
which can always be converted to or from \coin
by adjusting for current participation and inflation incurred thus far.

\end{document}